# EDUCATIONAL DATA MINING USING JMP

Sadiq Hussain[1] and Prof. G.C. Hazarika[2]

[1] System Administrator, Dibrugarh University, Dibrugarh Assam
[2] HoD, Mathematics, Dibrugarh University, Dibrugarh Assam

*ABSTRACT*

*Educational Data Mining is a growing trend in case of higher education. The quality of the Educational Institute may be enhanced through discovering hidden knowledge from the student databases/ data warehouses. Present paper is designed to carry out a comparative study with the TDC (Three Year Degree) Course students of different colleges affiliated to Dibrugarh University. The study is conducted with major subject wise, gender wise and category/caste wise. The experimental results may be visualized with Scatterplot3D, Bubble Plot, Fit Y by X, Run Chart, Control Chart etc. of the SAS JMP Software.*

## 1. INTRODUCTION

Dibrugarh University, the easternmost University of India was set up in 1965 under the provisions of the Dibrugarh University Act, 1965 enacted by the Assam Legislative Assembly. It is a teaching-cum-affiliating University with limited residential facilities. The University is situated at Rajabheta at a distance of about five kilometres to the south of the premier town of Dibrugarh in the eastern part of Assam as well as India. Dibrugarh, a commercially and industrially advanced town in the entire north-eastern region also enjoys a unique place in the fields of Art, Literature and Culture. The district of Dibrugarh is well known for its vast treasure of minerals (including oil and natural gas and coal), flora and fauna and largest concentration of tea plantations. The diverse tribes with their distinct dialects, customs, traditions and culture form a polychromatic ethnic mosaic, which becomes a paradise for the study of Anthropology and Sociology, besides art and culture. The Dibrugarh University Campus is well linked by roads, rails, air and waterways. The National Highway No.37 passes through the University Campus. The territorial jurisdiction of Dibrugarh University covers seven districts of Upper Assam, viz, Dibrugarh, Tinsukia, Sivasagar, Jorhat, Golaghat, Dhemaji and Lakhimpur. [1].

For educational data mining, the authors choose the Courses B.A. (Bachelor of Arts) and B.Sc. (Bachelor of Science) of Dibrugarh University. The authors collected the four years (2010-2014) digitized data from the Dibrugarh University Examination Branch. The authors tried to study and analyze the results of the candidates that are enrolled in the above mentioned courses gender-wise, caste-wise and subject-wise. There are lots of data mining tools and statistical models available. This paper gives emphasis on the mining tools and the statistical methods that would be best suited for knowledge discovery.

## 2. LITERATURE REVIEW

### 2.1 Data Mining

Data Mining is used to extract hidden patterns and information from the databases/ data warehouses. The data mining tools and algorithms may be used to predict future trends from the existing historical data. With the help of data mining tools, the large organizations may take knowledge driven decisions that were too time consuming earlier. It is like searching for pearls in the sea of data.

The Statisticians used manual approaches to provide business analysis and future trends for many years. At present the data mining tools and algorithms are not only used by the statisticians alone but also by the professionals from different spheres with more ease. So, the data miners able to answer the detailed questions quicker and may analyze the answer with the help of reports and graphical tools. [2].

Nowadays, as the storage and processing cost goes down, more and more organizations opt for data storage at much cheaper cost and at larger scale. The data may be extracted for information by classification, clustering, association rules, decision tree etc. [3].

### 2.2 Educational Data Mining

Data mining also called knowledge discovery in databases (KDD) is nowadays extensively used in the field of education. The term is coined as Educational Data Mining. Data Mining is an emerging field in education system to analyze the teaching learning environment and problems. [3]. The decision making becomes effective for the educational authorities with the help of data mining as they may identify and enhance the educational process by detecting the anomalies. [4]. Gabrilson et. al. [5] and Luan [6] used the data mining prediction techniques to determine the student performance on the basis of test score and group the students to find which students may finish their courses in time.

### 2.3 Jmp Software for Data Mining

SAS created JMP in 1989 to empower scientists and engineers to explore data visually. Since then, JMP has grown from a single product into a family of statistical discovery tools, each one tailored to meet specific needs. The software is visual, interactive, comprehensive and extensible. JMP Pro is the advanced analytics version of JMP that lets anyone use the data for better anticipation of the future and plan well for tomorrow.

Built with scientists and engineers in mind, JMP Pro statistical analysis software from SAS provides all the superior visual data access and manipulation, interactivity, comprehensive analyses and extensibility that are the hallmarks of JMP, plus a multitude of additional techniques. Our latest release adds advanced analytics like generalized regression, mixed-effects models, advanced consumer research analysis, reliability block diagrams and more. JMP Pro also features predictive modelling with cross-validation, model comparison and averaging features, exact tests and one-click bootstrapping.

More importantly, JMP Pro helps you construct a narrative and interactively communicate findings in ways others can readily understand and act upon. [7]

## 3. EXPERIMENTS AND EVALUATION

### 3.1 The Data Set

The authors had presented a small part of the Category, Subject and Gender based tables in Table 1, Table 2 and Table 3 for the analysis to be performed. The Examination Branch of Dibrugarh University provides various subject Codes for different subjects of BA and BSc courses.

The field 'Appeared' means the number of candidates appeared for that examination and 'Passed' means the number of candidates passed for that particular examination. The field 'PassPercentage' is the Passed Percentage of the Candidates for a particular course. The Category describes the General, SC,ST,OBC and MOBC students. The Gender describes the male and female candidates. The Subject describes the different Major subjects offered in BA and BSc Courses.

### 3.2 The Experiments and Evaluation

Statistical techniques including regression analysis, Bayesian Classifiers etc were used as a methodology. The data is collected from the Intel Xeon Based Server running SQL Server database. Only the required fields were extracted through SQL Query from the database. All the variables are defined in Tables I, II and III .

## 4. CONCLUSIONS AND FUTURE WORK

It may be concluded from the present study that the performances of the female candidates' are better than the male candidates' with reference to BA and BSc Examinations. Refer to the bubble plot in the figure 9. The results of the General category students are far better than the other category students. Refer to the Oneway analysis and Scatterplot 3D diagrams in the Figures 7 and 8. The analysis of major subjects in Arts Stream reveals that Bengali Major and Statistics Major Students excel in the examinations. The pass percentages for these subjects are 100%. The Education Major, Home Science Major and Sociology Major Students always score above average. The History Major, Mathematics Major and Sanskrit Major Students showed fluctuating trend. These may be visualized in the figures 1, 2 and 3 as depicted in the Run Chart, Control Chart and One way Analysis. With reference to the Run Chart, One way Analysis and Control IR Chart in the figures 4,5 and 6, for the Science Stream Subjects, it may be found that Geology Major, Electronics Major, Home Science Major and Statistics Major Students always excel and their performance is always above average.

In future the authors plan to develop a predictive model for the BA and BSc students based on various variables.

## ACKNOWLEDGMENTS

The authors express their gratitude to Prof. Alak Kr. Buragohain, Vice-Chancellor, Dibrugarh University for his inspiring words and allowing them to use the Examination data of the University. They generously thank Mr. N.A. Naik, Senior Programmer, Mumbai based firm for helping them to extract the .csv files from the SQL Server database. The authors would like to offer thanks to Prof. Jiten Hazarika Department of Statistics, Dibrugarh University for his valuable helps.

Table 1 : Sample Data for Year-wise Category-wise Course-wise Data of the B.A. and B.Sc. Candidates

| Year | Course | Gender | Appeared | Passed | Pass Percentage |
|---|---|---|---|---|---|
| 2010 | BA | M | 5285 | 4379 | 82.86 |
| 2010 | BA | F | 8160 | 6940 | 85.05 |
| 2011 | BA | M | 6691 | 5328 | 79.63 |
| 2011 | BA | F | 10088 | 8349 | 82.76 |

Table 2 : Sample Data for Year-wise Course-wise Subject-wise Data of the B.A. and B.Sc. Candidates

| Year | SubjectCode | SubjectName | Appeared | Passed | Pass Perecentage |
|---|---|---|---|---|---|
| 2010 | ANTM | MAJOR : ANTHROPOLOGY | 6 | 6 | 100 |
| 2010 | BOTM | MAJOR : BOTANY | 134 | 125 | 93.28 |
| 2010 | CHMM | MAJOR : CHEMISTRY | 121 | 110 | 90.91 |
| 2010 | ECOM | MAJOR : ECONOMICS | 1 | 1 | 100 |

Table 3 : Sample Data for Year-wise Course-wise Gender-wise Data of the B.A. and B.Sc. Candidates

| Year | Course | Caste | Appeared | Passed | PassPercentage |
|---|---|---|---|---|---|
| 2010 | BSC | General | 266 | 244 | 91.73 |
| 2010 | BSC | MOBC | 70 | 61 | 87.14 |
| 2010 | BSC | OBC | 351 | 310 | 88.32 |
| 2010 | BSC | SC | 51 | 43 | 84.31 |
| 2010 | BSC | ST | 80 | 70 | 87.5 |
| 2011 | BSC | General | 354 | 314 | 88.7 |

**Run Chart of Pass Percentage**

Fig 1: Run Chart of the different BA Major Subjects w.r.t. pass percentage

Fig 2: Oneway Analysis Fit Y By X of the different BA Major Subjects w.r.t. pass percentage

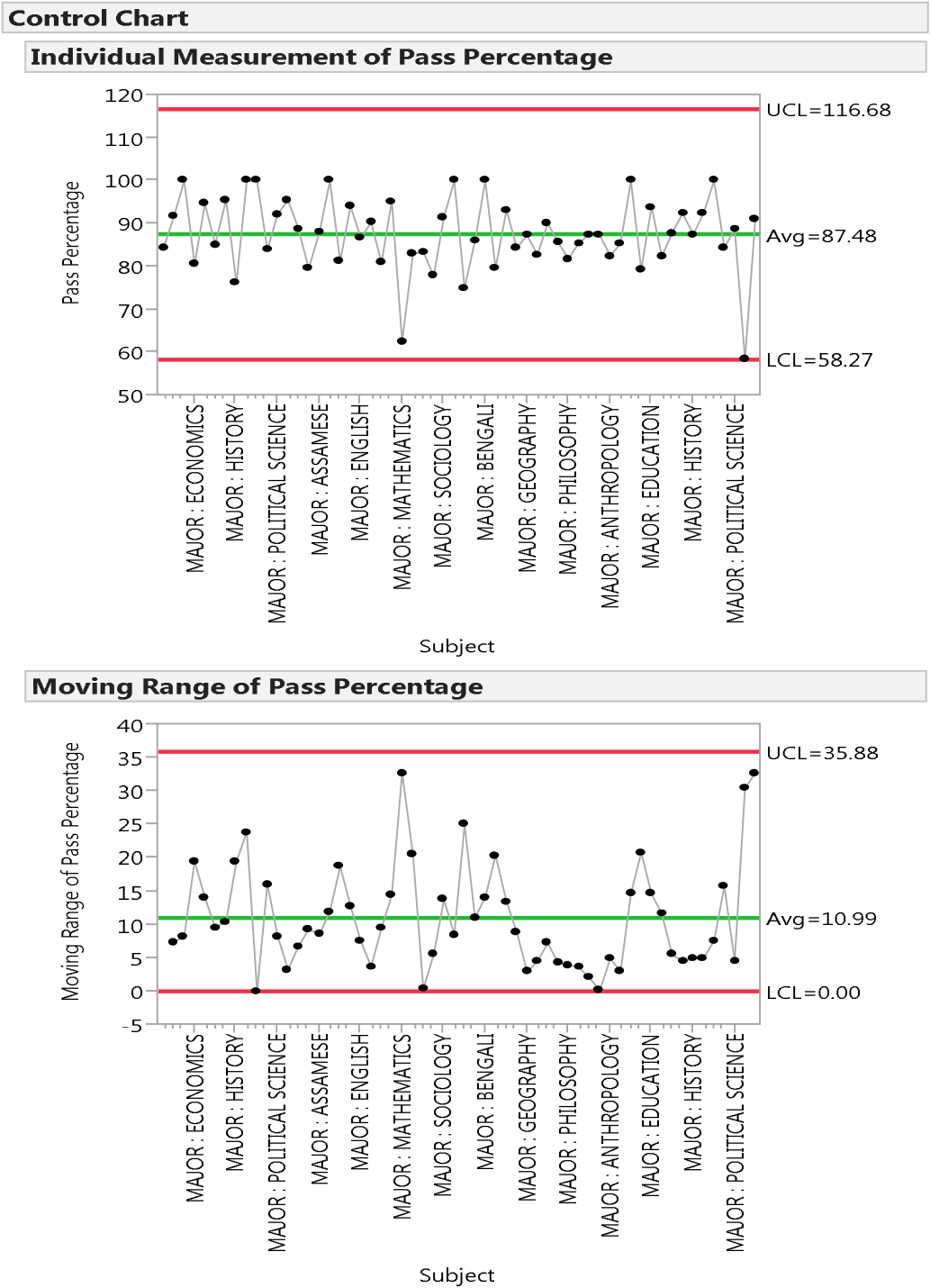

Fig 3: Control Chart of the different BA Major Subjects w.r.t. pass percentage

**Run Chart of Pass Percentage**

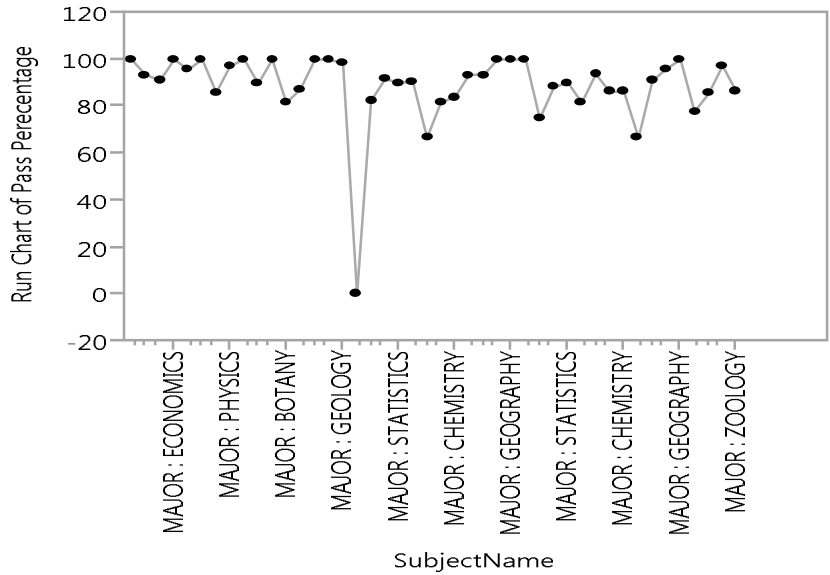

Fig 4: Run Chart of the different BSc Major Subjects w.r.t. pass percentage

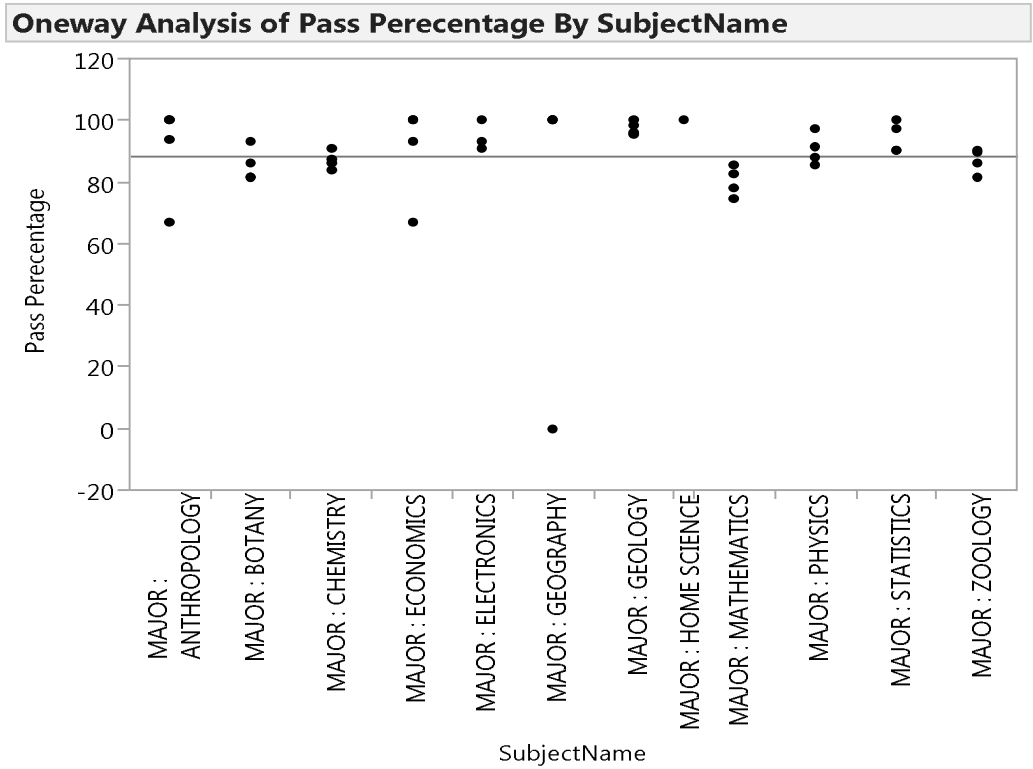

Fig 5: Oneway Analysis Fit Y By X of the different BSc Major Subjects w.r.t. pass percentage

## Control Chart

### Individual Measurement of Pass Percentage

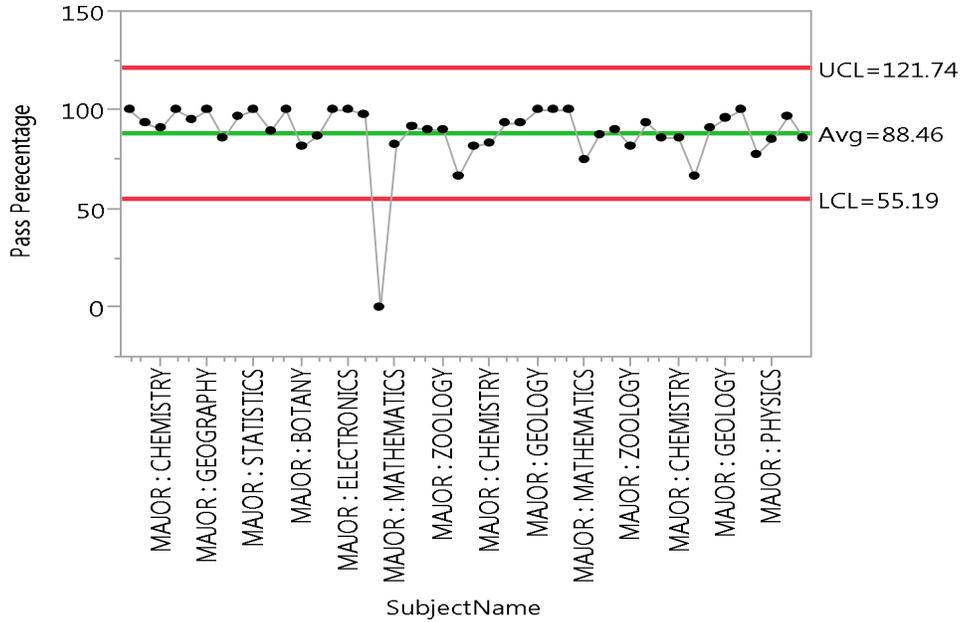

### Moving Range of Pass Percentage

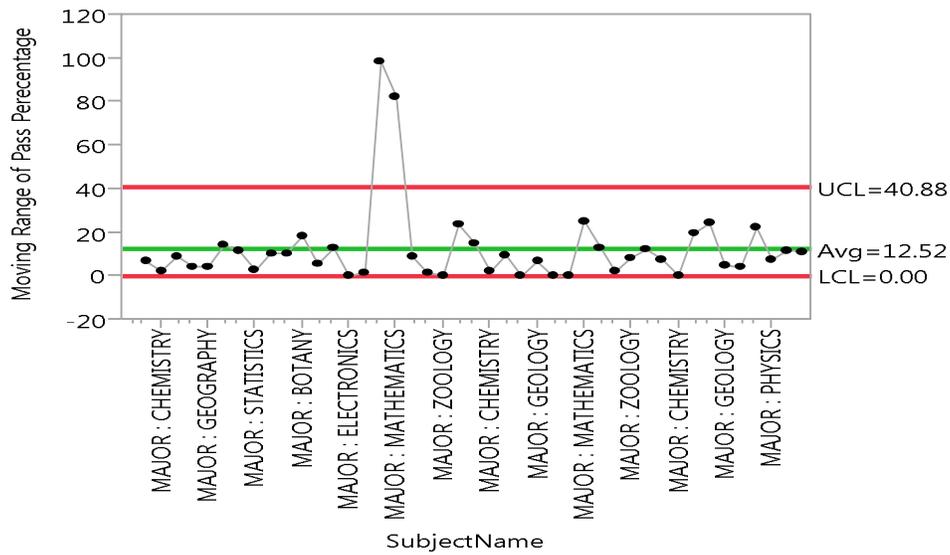

Fig 6: Control Chart of the different BSc Major Subjects w.r.t. pass percentage

**Oneway Analysis of Pass Percentage By Caste**

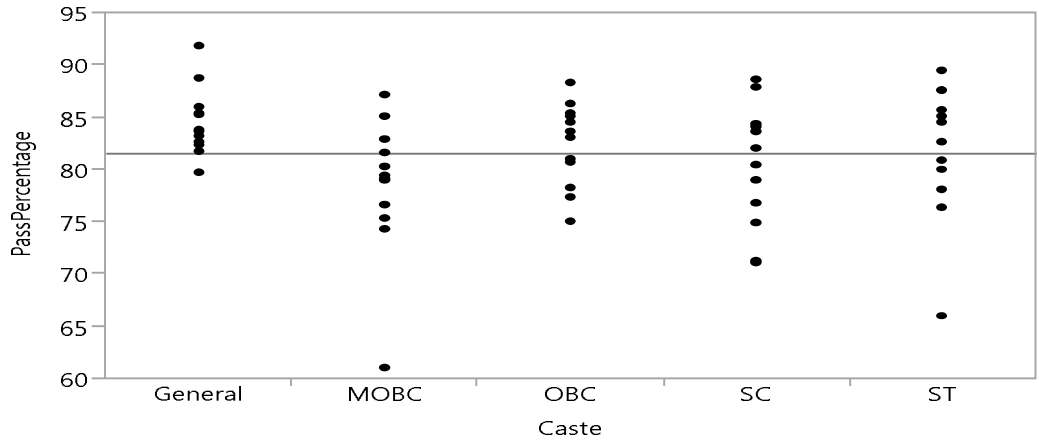

Fig 7: Oneway Analysis Fit Y By X of the different category/caste students of BA & BSc course w.r.t. pass percentage

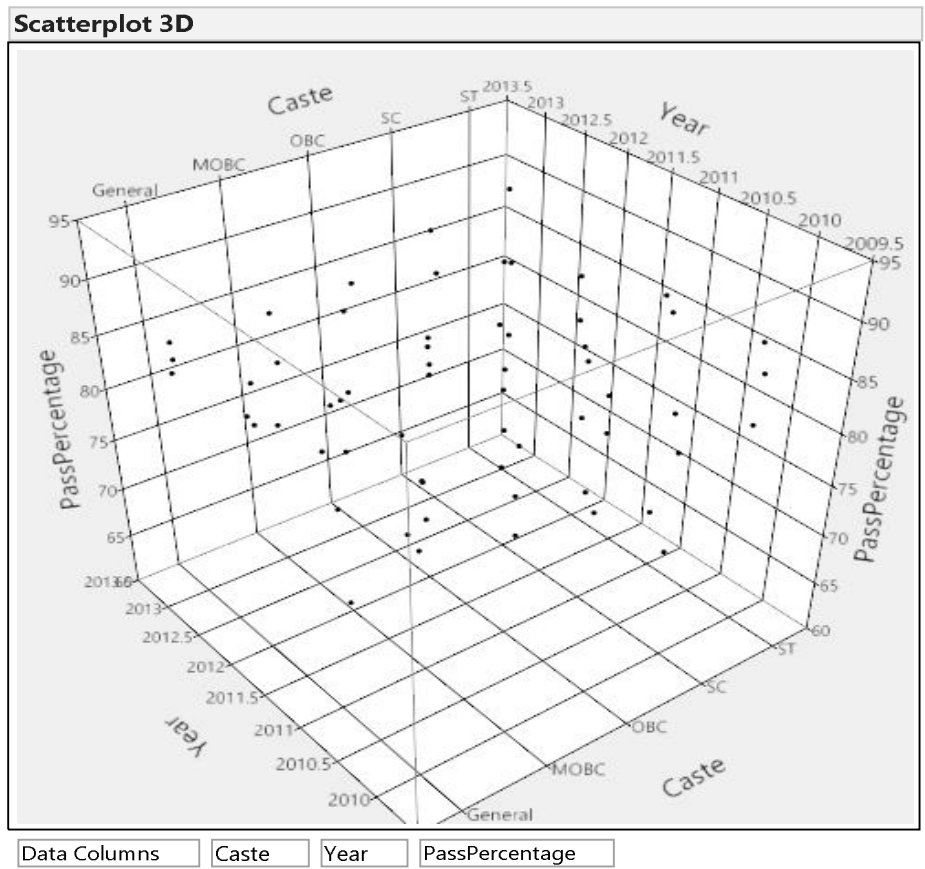

Fig 8: Scatterplot3D of the different category/caste students of BA & BSc course w.r.t. pass percentage

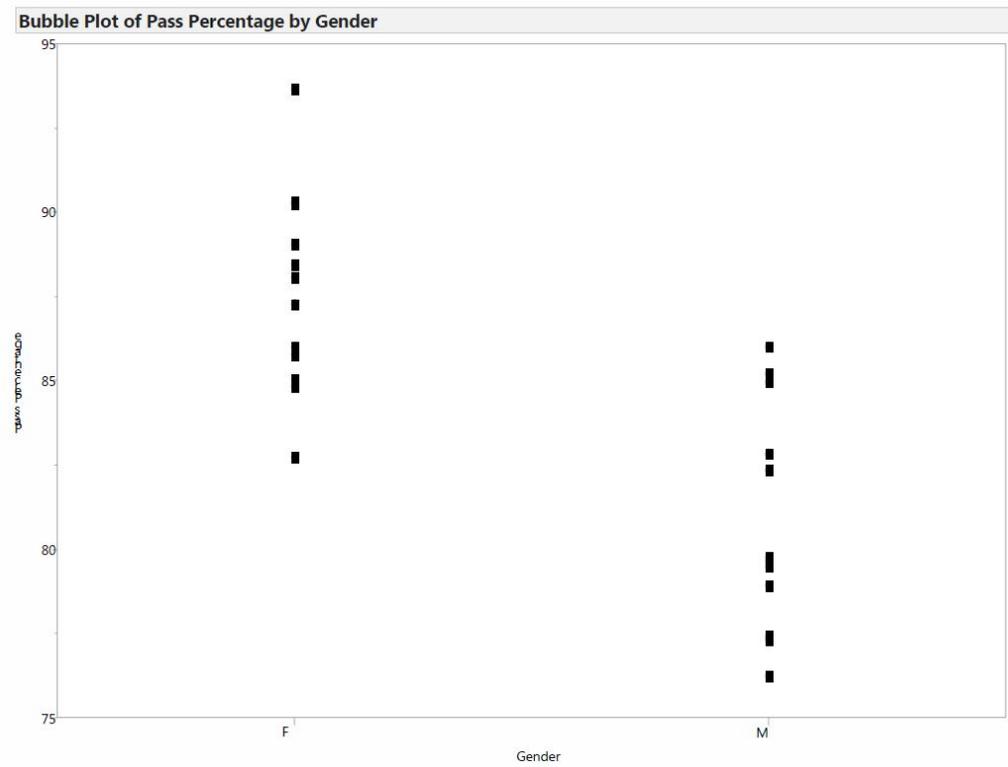

Fig 9: Bubble Plot of BA & BSc students gender-wise w.r.t. pass percentage